\documentclass[12pt]{article}
 \usepackage[dvips,final]{graphicx}
  \usepackage{amsmath}
   \usepackage{amssymb}
    \usepackage{pifont}

\begin{document}

\textbf{\large Pion distribution amplitude -- from theory to data
         (CELLO, CLEO, E-791, JLab F(pi))}
\vspace*{6mm}

Alexander~P.~Bakulev\footnote{Talk presented at 
the 13th International Seminar on High Energy Physics 
``Quarks-2004'',
Pushkinskie Gory, Russia, May 24-30, 2004. 
It is based on results obtained in collaboration with 
S.~Mikhailov, K.~Passek-Kumeri\v{c}ki, W.~Schroers, N.~G.~Stefanis}

\textit{\small Bogoliubov Laboratory of Theoretical Physics,
               JINR, Dubna, 141980 Russia}
\vspace*{6mm}

\textbf{Outline}: 
\begin{itemize}
\item  What is the pion distribution amplitude ${\varphi_{\pi}(x)}$? 
\item  Nonperturbative part: How to obtain ${\varphi_{\pi}(x)}$ from QCD sum rules;
\item  Perturbative part: NLO light-cone sum rules  $\Rightarrow$
       CLEO experiment on ${F^{\gamma\gamma^*\pi}(Q^2)}\Rightarrow$ 
       constraints on ${\varphi_{\pi}(x)}$ and $\lambda_q^2$;  
\item  Perturbative addition: Diffractive dijet production (E791 data);
\item  Perturbative addition: Pion electromagnetic form factor (CEBAF data);
\item  Conclusions.
\end{itemize}

\textbf{The main object of this talk is the pion distribution amplitude} (DA),
which can be defined through the matrix element 
of a nonlocal axial current on the light cone
\begin{eqnarray} 
 \langle{0\mid \bar d(z)\gamma_{\mu}\gamma_5 E(z,0) u(0)\mid \pi(P)}\rangle\Big|_{z^2=0}
 &=& i f_{\pi}P_{\mu}
    \int_{0}^{1} dx\ e^{ix(zP)}\
     \varphi_{\pi}(x,\mu^2)\,,
    \label{eq:PiME}
\end{eqnarray}
which is explicitly  gauge-invariant
due to the presence of the Fock--Schwinger connector
$E(z,0) = {\cal P}e^{i g \int_0^z A_\mu(\tau) d\tau^\mu}$. 
The physical meaning of this object is quite transparent:\\
\begin{minipage}{0.45\textwidth}
It is the amplitude for the transition of the physical pion $\pi(P)$ 
to a pair of valence quarks $u$ and $d$, separated at light-cone
(see graphical image to the right),
with momentum fractions $xP$ and $\bar{x}P$,
correspondingly (here $\bar{x}\equiv 1-x$).
\end{minipage}\hspace{0.1\textwidth}
\begin{minipage}[h]{0.45\textwidth}
\vspace*{-0.5mm}
\noindent\hspace*{-2mm}\includegraphics[width=\textwidth]{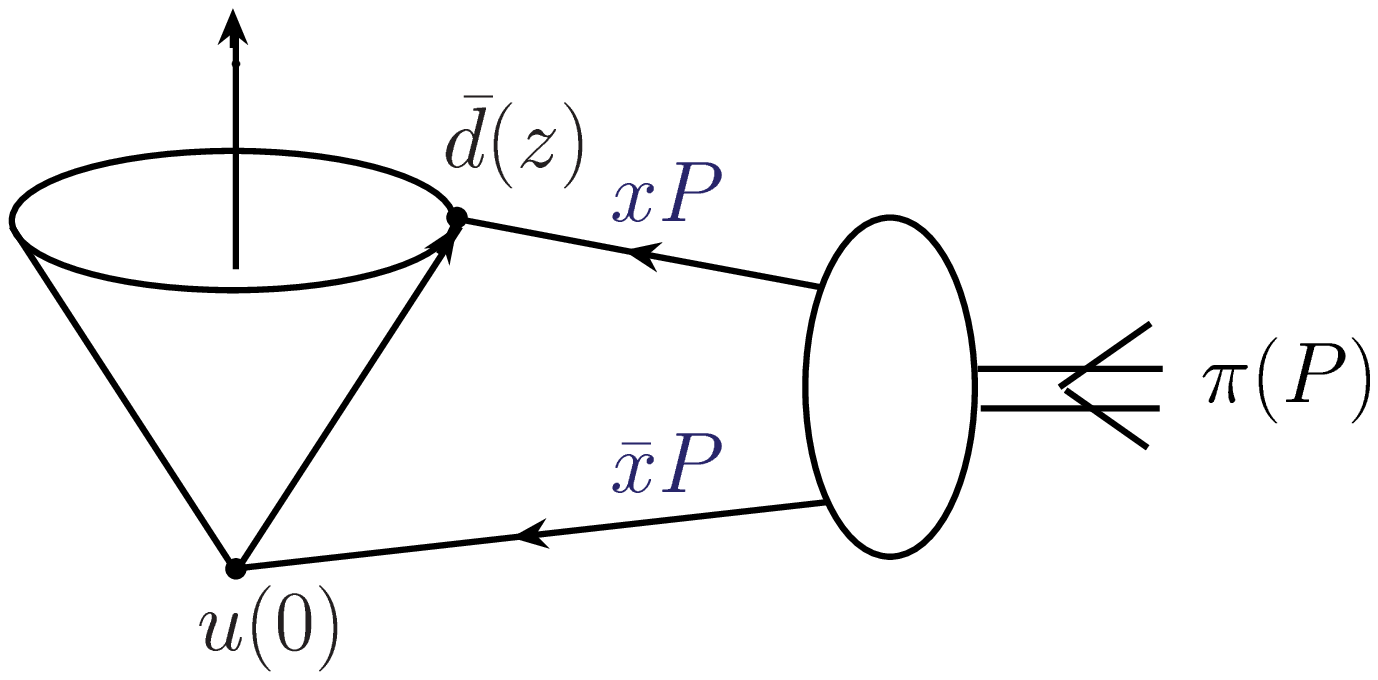}
\vspace*{2mm}
\end{minipage}
This object inevitably appears in applying perturbative QCD to hard processes
with pions in the initial or the final state
as a result of QCD factorization theorems~\cite{CZ77,ER80,LB79}
and it includes nonperturbative information 
about the physical pion.
It has the following properties:
\begin{itemize}
 \item normalized to unity $\int_{0}^{1} dx\,\varphi_{\pi}(x,\mu^2)=1$;
 \item $x\rightleftarrows\bar{x}$ symmetric: $\varphi_{\pi}(x,\mu^2)=\varphi_{\pi}(\bar{x},\mu^2)$;
 \item obeys the ER-BL evolution equation~\cite{ER80,LB79}
  with respect to $\mu^2$;
 \item in the 1-loop approximation 
 $\varphi_\pi(x;\mu^2\to\infty) = \varphi^\text{as}(x) = 6x (1-x)$.
\end{itemize}

It is convenient to represent the pion DA as an expansion in terms 
of Gegenbauer polynomials $C^{3/2}_n(2x-1)$,
being the 1-loop eigenfunctions of the ER-BL kernel:
\begin{eqnarray}
 \varphi_\pi(x;\mu^2)
  &=& \varphi^\text{as}(x)
      \Bigl[ 1 + a_2(\mu^2) C^{3/2}_2(2x-1)
               + a_4(\mu^2) C^{3/2}_4(2x-1) 
               + \text{...}
      \Bigr]\,.
\end{eqnarray}
That means to transfer all the $\mu^2$-dependence of the pion DA
into the Gegenbauer coefficients $\left\{ a_2(\mu^2), a_4(\mu^2), \ldots \right\}$.
This scheme can be effectively applied at the 2-loop level as well~\cite{MR86ev,Mul94}.

\begin{figure}[hb]
$$\includegraphics[width=\textwidth]{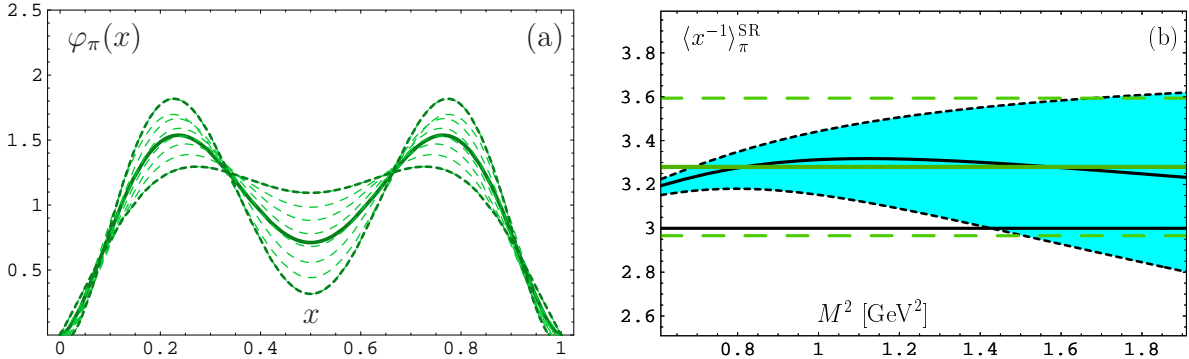}$$\vspace*{-9mm}

   \caption{\label{fig:b410g}\footnotesize
    (a) The bunch of pion DAs extracted from NLC QCD sum rules.
     Parameters of the bold-faced curve are $a_2^\textbf{b.f.}=+ 0.188$ 
     and $a_4^\textbf{b.f.}= - 0.130$.
    (b) The results for the inverse moment $\langle{x^{-1}}\rangle_\pi$ 
    as a function of the Borel parameter $M^2$
    obtained using a special model-independent sum rule.
    The shaded area corresponds to the 10\%-variation 
    of the threshold parameter $s_0$. Dashed straight lines 
    show the allowed window for $\langle{x^{-1}}\rangle_\pi^\text{SR}$.}
\end{figure}
\textbf{In order to obtain the pion DA in the theory},
one is obliged to use some nonperturbative approach.
Historically, the first nontrivial model 
has been constructed by Chernyak and Zhitnitsky (CZ)~\cite{CZ82}
using the standard QCD sum rule approach and estimating the first two moments
of the pion DA: $\langle{\xi^2}\rangle_\pi$ and $\langle{\xi^4}\rangle_\pi$.
After that, Mikhailov and Radyushkin 
realized 
that in doing so CZ highly overestimated these moments
and suggested to use instead the non-local condensate (NLC) approach~\cite{MR86}.
We have used the NLC QCD sum rules and obtained the first five moments
of the pion DA, $\langle{\xi^{2N}}\rangle_\pi$ with $N=1,...,5$.
Just for illustration, we present here the simplest scalar condensate 
of the used NLC model:
\begin{eqnarray}
 \label{eq:ScaNLC}
  \langle{\bar{q}(0)q(z)}\rangle 
   &=& \langle{\bar{q}(0)q(0)}\rangle\, e^{-|z^2|\lambda_q^2/8}\,.
\end{eqnarray}
This model is determined by a single scale parameter $\lambda_q^2 = \langle{k^2}\rangle$
characterizing the average momentum of quarks in the QCD vacuum.
It has been estimated in QCD SRs and on the lattice:
\begin{eqnarray}\label{eq:Lam_q^2}
 \lambda_q^2
  &=&\left\{
      \begin{array}{lc}
       0.4\pm 0.1\text{~GeV}^2 & \left[\text{\ QCD SRs~\cite{BI82}\ }\right]\\
       0.5\pm0.05\text{~GeV}^2 & \left[\text{\ QCD SRs~\cite{OPiv88}\ }\right]\\
       \approx 0.4-0.5\text{~GeV}^2 & \left[\text{\ Lattice~\cite{DDM99,BM02}\ } \right]
      \end{array} 
     \right.
\end{eqnarray}
NLC sum rules for the pion DA produce~\cite{BMS01} 
a ``\textbf{bunch}'' of self-consistent 2-parameter models at $\mu^2\simeq 1$ GeV$^2$:
\begin{eqnarray}
 \label{eq:2p-Bunch}
  \varphi_\pi(x)
  &=& \varphi^{\text{as}}(x)
      \left[1 + a_2 C^{3/2}_2(2x-1) + a_4 C^{3/2}_4(2x-1)
      \right]\,.
\end{eqnarray}
For the most favorite value of the vacuum nonlocality parameter $\lambda_q^2=0.4$ GeV$^2$
we have the bunch of pion DAs presented in Fig.\,\ref{fig:b410g}a.
By self-consistency we mean that the value of the inverse moment
for the whole bunch $\langle{x^{-1}}\rangle_\pi^{\text{bunch}} = 3.17\pm0.10$
is in agreement with the independent estimation from the special sum rule,
$\langle{x^{-1}}\rangle_\pi^{\text{SR}}=3.30\pm0.30$, see Fig.\,\ref{fig:b410g}b.

\begin{figure}[t]
 $$\includegraphics[width=\textwidth]{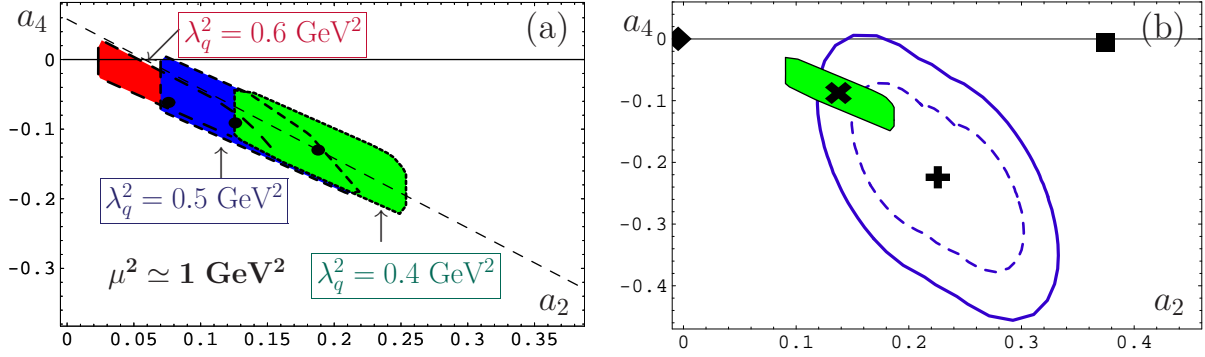}$$\vspace*{-9mm}
 
   \caption{\label{fig:b456}\footnotesize
    (a) The bunches of pion DAs extracted from NLC QCD sum rules
    in the $(a_2,a_4)$-plane for three values of the nonlocality parameter $\lambda_q^2$.
    (b) Comparison of the NLC-bunch evolved to $\mu^2=5.76$~GeV$^2$ 
    with the CLEO data constraints for $\lambda_q^2 = 0.4\text{ GeV}^2$. 
    The $1\sigma$- and the $2\sigma$-contours are shown in dashed and solid lines, correspondingly.}
\end{figure}

We also extract the corresponding bunches for two other values 
of $\lambda_q^2=0.5$~GeV$^2$ and $\lambda_q^2=0.6$~GeV$^2$, 
and show the results as allowed regions in the $(a_2,a_4)$-plane 
in Fig.\,\ref{fig:b456}a.

\textbf{NLO light-cone sum rules (LCSR) and the CLEO data on $F_{\gamma\gamma^*\pi}(Q^2)$}
allow one to obtain constraints on $\varphi_\pi(x)$ directly from the experimental data.
A natural question arises: Why does one need to use LCSRs?
The answer is that for $Q^2\gg m_\rho^2$, $q^2\ll m_\rho^2$
pQCD factorization is valid only in leading twist
but higher twists are also of importance~\cite{RR96}.
The reason is quite evident: if $q^2\to 0$ one needs to take into account 
the interaction of a real photon at long distances of order of $O(1/\sqrt{q^2})$.
To account for long-distance effects in perturbative QCD, 
one needs to introduce a light-cone DA of a real photon.
In the absence of reliable information about the photon DA,
Khodjamirian~\cite{Kho99} suggested to use the LCSR approach,
which effectively accounts for long-distances effects of a real photon,
using the quark-hadron duality in the vector channel 
and a dispersion relation in $q^2$:
$$F_{\gamma\gamma^*\pi}(Q^2, 0) 
  =
   \frac{1}{\pi}\int_{0}^{s_0}
    \frac{\textbf{Im}F_{\gamma^*\gamma^*\pi}^\text{PT}(Q^2,-s)}{m_\rho^2}
     \exp\left[\frac{m_\rho^2-s}{M^2}\right]ds
  + \frac{1}{\pi}\int_{s_0}^{\infty}
    \frac{\textbf{Im}F_{\gamma^*\gamma^*\pi}^\text{PT}(Q^2,-s)}{s}ds\,,
$$
with $s_0\simeq1.5$~GeV$^2$ -- effective threshold in vector channel,
 $M^2$ -- Borel parameter ($0.5-0.9$ GeV$^2$).
We revised the NLO LCSR approach of~\cite{SY99} in performing the CLEO data analysis
along the following lines~\cite{BMS02}:
\begin{itemize}
\item  An accurate NLO evolution for both $\varphi(x, Q^2_\text{exp})$ 
       and $\alpha_s(Q^2_\text{exp})$,
       taking into account heavy quark thresholds.
\item The relation  between the ``nonlocality" scale and 
      the the twist-4 magnitude $\delta^2_\text{Tw-4} \approx \lambda_q^2/2$
      was used to  re-estimate 
      $\delta^2_\text{Tw-4}= 0.19 \pm 0.02$ at $\lambda_q^2=0.4$ GeV$^2$.
\item Constraints on $\langle{x^{-1}}\rangle_\pi$ from the CLEO data.         
\end{itemize}
As a result, we have obtained reasonable agreement of our bunch with the CLEO data 
for $\lambda_q^2 = 0.4\text{ GeV}^2$, see Fig.\,\ref{fig:b456}b
(with (\ding{117}) = {asymptotic} DA, (\ding{54}) = BMS model,
 ({\footnotesize \ding{110}}) = CZ DA, and (\ding{58}) corresponds to the best-fit point).

\begin{figure}[t]
 $$\includegraphics[width=\textwidth]{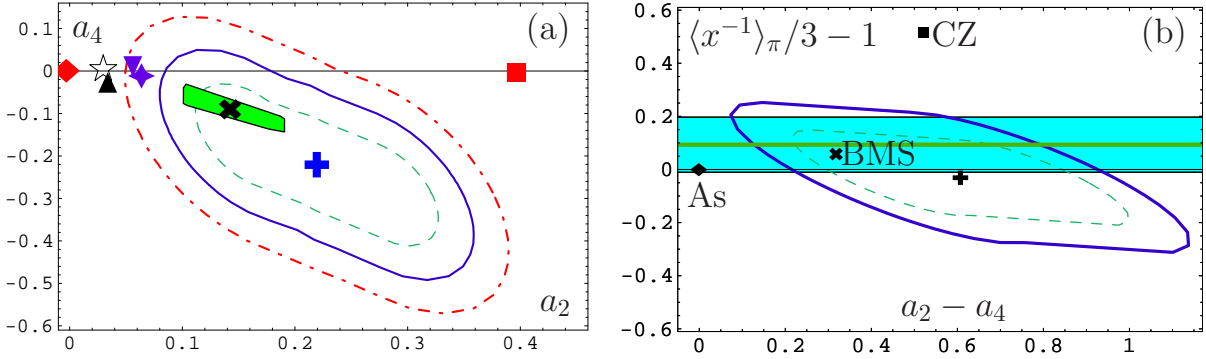}$$\vspace*{-9mm}
 
   \caption{\label{fig:CLEO20_INV}\footnotesize
    (a) Comparison of the NLC-bunch evolved to $\mu^2=5.76$~GeV$^2$ with the CLEO data constraints
    for $\lambda_q^2 = 0.4\text{ GeV}^2$. The $1\sigma$-, $2\sigma$- and $3\sigma$-contours are shown
    as dashed, solid and dash-dotted contours. For details see in the text.
    (b) Comparison of BMS, CZ and asymptotic DAs at the QCD sum-rule scale $\mu^2\simeq1$~GeV$^2$ 
    with the CLEO data constraints for $\lambda_q^2 = 0.4\text{ GeV}^2$ 
    in terms of rotated axes $a_2-a_4$ and $a_2+a_4$.
    The $1\sigma$- and $2\sigma$-contours are shown as dashed and solid lines.}
\end{figure}

In order to make our conclusions more valuable,
we have adopted a 20\% uncertainty in the magnitude of the twist-4 contribution,
$\delta_\text{Tw-4}^2 = 0.19\pm0.04$~GeV$^2$, 
and produced new $1\sigma$-, $2\sigma$- and $3\sigma$-contours 
dictated by the CLEO data~\cite{BMS03},
see Fig.\,\ref{fig:CLEO20_INV}a
in parallel with available 2-Gegenbauer models:
asymptotic DA, BMS model, CZ DA 
(they are shown in the same manner as in Fig.\,\ref{fig:b456}b),
three instanton-based models, viz., (\ding{73})~\cite{PPRWG99},
{\footnotesize\ding{115}}~\cite{ADT00}, 
and (\ding{70}) 
(using in this latter case $m_q=325$~MeV, $n=2$, and $\Lambda=1$~GeV)~\cite{Pra01},
and a recent transverse lattice result ({\footnotesize\ding{116}})~\cite{Dal02}.
We see that even with a 20\% uncertainty in the twist-4, the CZ DA is excluded 
\textbf{at least} at the \textbf{$4\sigma$}-level, 
whereas the asymptotic DA -- at the \textbf{$3\sigma$}-level.
Our bunch is mainly inside the $1\sigma$-region 
and other nonperturbative models are near the 3$\sigma$-boundary.

We also plot the CLEO data in the plane $(X,Y)$
with $X=a_2-a_4$ and $Y=a_2+a_4=\langle{x^{-1}}\rangle_\pi/3-1$,
where the Gegenbauer coefficients $a_2$ and $a_4$ refer
to the NLC sum-rule scale $\mu^2\simeq1$~GeV$^2$.
The result is shown in Fig.\,\ref{fig:CLEO20_INV}b,
where the comparison of the CLEO data constraints 
directly with the model-independent bound 
$\frac{1}{3}\langle x^{-1} \rangle^\text{SR}_{\pi} - 1 = 0.1\pm0.1$
from the NLC QCD sum rule (shaded strip in figure)
is done.
Again we see a good agreement of a theoretical ``tool'' 
of different origin 
with the CLEO data.
Here, we should also mention other estimations 
of the pion DA inverse moment.
Bijnens\&Khodjamirian produced an estimate
$\frac{1}{3}\langle x^{-1} \rangle_{\pi} - 1 = 0.24 \pm 0.16$
using data on the pion electromagnetic form factor in the LCSR approach~\cite{BiKho02},
whereas Ruiz Arriola\&Broniowski
obtained in their model of the pion DA with an infinite number of Gegenbauer harmonics
the result $\frac{1}{3}\langle x^{-1} \rangle_{\pi} - 1 = 0.25\pm 0.1$~\cite{ArriBro02}.

\begin{figure}[h]
 $$\includegraphics[width=\textwidth]{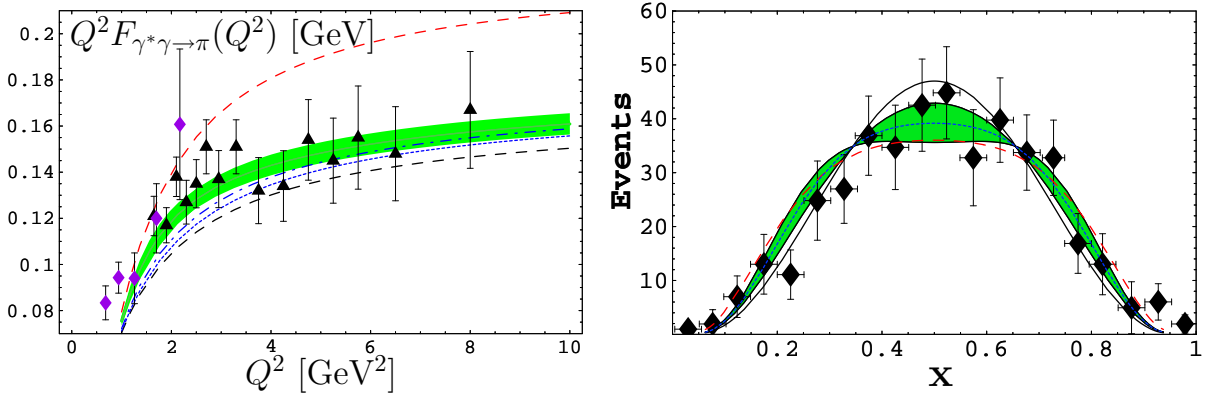}$$\vspace*{-9mm}
 
   \caption{\label{fig:CLEO_FF}\footnotesize
    (a) $\gamma^*\gamma\to\pi$ Transition form factor in comparison 
    with the CELLO ({\tiny\ding{117}})~\cite{CELLO91} and 
    the CLEO ({\tiny\ding{115}})~\cite{CLEO98} data. 
    For details see in the text.
    (b)  Comparison of the asymptotic DA (solid line), CZ DA (dashed line), 
    and the  BMS ``bunch'' of pion DAs (strip) with the E791 data (\ding{117})~\protect\cite{E79102}.}
\end{figure}
To finish our discussion about the CLEO data constraints in the NLO LCSR approach,
we show in Fig.\,\ref{fig:CLEO_FF}a 
the plot of $Q^2F_{\gamma^*\gamma\to\pi}(Q^2)$
for our bunch (shaded strip), CZ DA (upper dashed line), asymptotic DA (lower dashed line),
and two instanton-based models (dotted~\cite{PPRWG99} and dash-dotted~\cite{PR01} lines)
in comparison with the CELLO and the CLEO data.
We see that the BMS bunch describes rather well all data for $Q^2\gtrsim1.5$ GeV$^2$.

\textbf{Diffractive Dijet Production:} What can add the E791 data to our analysis?
The diffractive dijet production in $\pi+A$ collisions has been suggested as a tool 
to extract the profile of the pion DA by Frankfurt et al. in 1993~\cite{FMS93}.
They argued that the jet distribution 
with respect to  the longitudinal momentum fraction
has to follow the quark momentum distribution in the pion 
and hence provides a direct measurement of the pion DA.
As it was shown just recently in~\cite{BISS02} (see also~\cite{NSS01}),
beyond the leading logarithms in energy this proportionality does not hold.
Braun et al. found that the longitudinal momentum fraction distribution 
of the jets for the non-factorizable contribution 
turns out to be the same as for the factorizable contribution
with the asymptotic pion distribution amplitude. 
We have used this convolution approach of Braun et al.
to estimate the distribution of jets in this experiment
for our bunch of pion DAs in comparison with the asymptotic and the CZ DAs~\cite{BMS03}.
Results are shown in Fig.\,\ref{fig:CLEO_FF}b.
It is interesting to note that the corresponding $\chi^2$ values are:
as -- 12.56; CZ -- 14.15; BMS -- 10.96 (accounting for 18 data points).
The main conclusion from this comparison: 
\textbf{all three DAs are compatible with the E791 data}.
Hence, this experiment cannot serve as a safe profile indicator.

\begin{figure}[h]
 $$\includegraphics[width=0.5\textwidth]{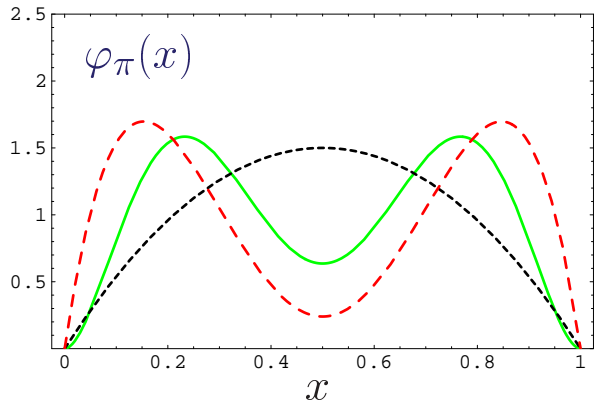}$$\vspace*{-9mm}
 
   \caption{\label{fig:BMS_CZ_DAs}\footnotesize
    Comparison of the asymptotic (dotted line), the CZ (dashed line),
    and the BMS DAs (solid line).}
\end{figure}
Let us say a few words about similarities and differences 
between the CZ and BMS DAs. 
Both are two-humped, 
but the CZ DA is strongly end-point enhanced, 
whereas the BMS DA is end-point suppressed!
And the reason for this behaviour is physically evident:
nonlocal quark condensate reduces pion DA in the small $x$ region
and enhances in the vicinity of the point $x\simeq0.2$.
In order to keep the norm equal to unity,
it is forced to have in the central region some reduction as well.

\textbf{Pion electromagnetic form factor}: How well is the BMS bunch 
in comparison with the JLab data on the pion form factor?
We have calculated the pion form factor in analytic NLO pQCD~\cite{BPSS04}
\begin{eqnarray}
  F_{\pi}(Q^{2};\mu_\text{R}^{2})
  =  F_{\pi}^\text{LD}(Q^{2})
  +  F_{\pi}^\text{Fact-WI}(Q^2;\mu_\text{R}^{2})\,,
 \label{eq:Q2Pff}
\end{eqnarray}
with taking into account the soft part $F_{\pi}^\text{LD}(Q^{2})$
via the local duality approach
and the factorized contribution
\begin{eqnarray}
 \label{eq:Fpi-Mod}
  F_{\pi}^\text{Fact-WI}(Q^2;\mu_\text{R}^{2})
  &=& \left(\frac{Q^2}{2s_0^\text{2-loop}+Q^2}\right)^2
       F_{\pi}^\text{Fact}(Q^2;\mu_\text{R}^{2})
\end{eqnarray}
has been corrected via a power-behaved pre-factor
(with $s_0^\text{2-loop}\approx0.6$~GeV$^2$)
in order to respect the Ward identity 
at $Q^2=0$ and preserve its high-$Q^2$ asymptotics.
In our analysis $F_{\pi}^\text{Fact}(Q^2;\mu_\text{R}^{2})$ 
has been computed to NLO~\cite{DR81,MNP99a}, 
using Analytic Perturbation Theory~\cite{SS97,DVS0012,SSK99} 
and trading the running coupling and its powers for analytic expressions 
in a non-power series expansion,
i.e.,
\begin{eqnarray}
 \left[F_{\pi}^\text{Fact}(Q^2; \mu_\text{R}^{2})\right]_\text{MaxAn}
  \ =\ \bar{\alpha}_\text{s}^{(2)}(\mu_\text{R}^{2})\, {\cal F}_{\pi}^\text{LO}(Q^2)
   + \frac{1}{\pi}\,
      {\cal A}_{2}^{(2)}(\mu_\text{R}^{2})\,
       {\cal F}_{\pi}^\text{NLO}(Q^2;\mu_\text{R}^{2})\,,
\label{eq:pffMaxAn}
\end{eqnarray}
with $\bar{\alpha}_\text{s}^{(2)}$ and ${\cal A}_{2}^{(2)}(\mu_\text{R}^{2})$ 
being the 2-loop analytic images of $\alpha_\text{s}(Q^2)$ 
and $\left(\alpha_\text{s}(Q^2)\right)^2$,
correspondingly (see~\cite{BPSS04} for more details), 
whereas
${\cal F}_{\pi}^\text{LO}(Q^2)$ and 
${\cal F}_{\pi}^\text{NLO}(Q^2;\mu_\text{R}^{2})$ 
are the LO and NLO parts of the factorized form factor,
respectively. 
\begin{figure}[h]
 $$\includegraphics[width=\textwidth]{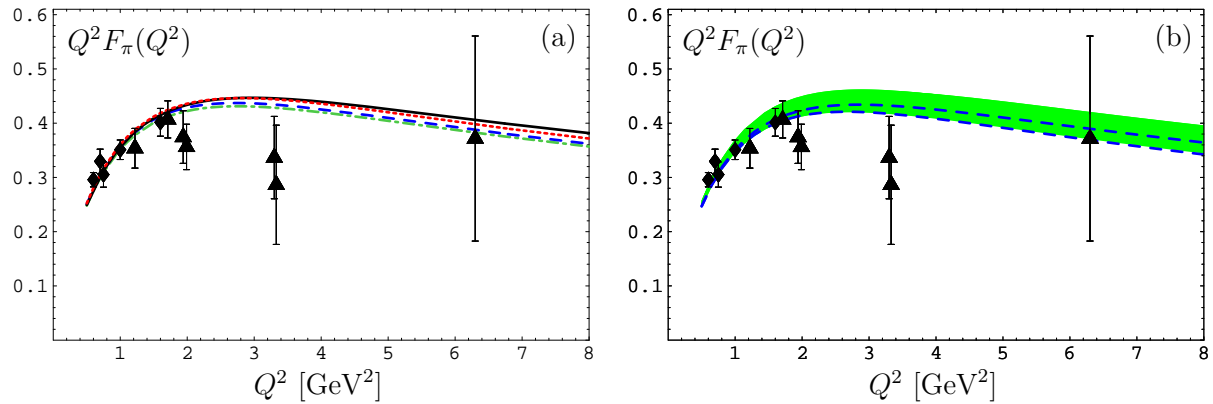}$$
 \vspace*{-8mm}
 
 \caption{\label{fig:pidatasum}\footnotesize
         Pion electromagnetic form factor in comparison 
         with the JLab ({\tiny\ding{117}})~\protect\cite{JLab00} 
         and Bebek et al. ({\tiny\ding{115}})~\protect\cite{FFPI73} data. 
     (a) The solid line represents results obtained with $\mu^2_\text{R}=1~\text{GeV}^2$,
         the dashed line -- with $\mu^2_\text{R}=Q^2$, 
         the dotted line -- with the BLM scale, and
         the dash-dotted line --- in the $\alpha_V$-scheme.
     (b) Predictions based on the BMS ``bunch'' of pion DAs (strip) and
         the asymptotic DA (dashed lines).
         The green strip contains the NLC QCD sum-rule uncertainties (due to the BMS bunch)
         and scale-setting ambiguities at the NLO level 
         (in the case of the asymptotic DA these ambiguities are represented 
          by two dashed lines).}
\end{figure}
The result of this analysis is presented in Fig.\,\ref{fig:pidatasum},
where we show $F_{\pi}(Q^{2})$ for the BMS ``bunch'' and using the
``Maximally Analytic'' procedure, 
which improves the previously introduced~\cite{SSK99} ``Naive Analytic'' one. 
The new procedure with the analytic running coupling and analytic versions of its powers
gives us practical independence of the scheme/scale setting 
(see Fig.\,\ref{fig:pidatasum}a and the figure caption for details)
and provides results in a rather good agreement with the
experimental data \cite{FFPI73,JLab00}. 
We see that the form-factor predictions are only slightly larger 
than those resulting when using the asymptotic DA (see Fig.\,\ref{fig:pidatasum}b).
\bigskip

\textbf{Conclusions}.
\begin{itemize}
 \item  The QCD sum rule method with NLCs for the pion DA 
  gives us admissible sets (bunches) of DAs for each {$\lambda_q$} value.

\item The NLO LCSR method produces new constraints on the pion DA parameters 
 ({$a_2,a_4$}) in conjunction with the CLEO data.

\item Comparing NLC sum-rule results with the new CLEO constraints
 allows us to fix the value of QCD vacuum nonlocality $\lambda_q^2=0.4$~GeV$^2$.

\item The corresponding bunch of pion DAs {agrees well} with the E791 data 
 on diffractive dijet production and with the JLab F(pi) data 
 on the pion electromagnetic form factor. 

\item Analytic perturbation theory with non-power NLO for the pion form factor 
 {diminishes scale-setting ambiguities} already at the NLO level, 
  rendering still higher-order corrections virtually superfluous. 
\end{itemize}
\bigskip

\textbf{Acknowledgments:} I wish to thank Prof.\ Klaus Goeke
for the warm hospitality at Bochum University,
where the major part of this investigation was carried out.
This work was supported in part by the Deutsche Forschungsgemeinschaft
(Projects 436 KRO 113/6/0-1 and 436 RUS 113/752/0-1),
the Heisenberg--Landau Programme,
the COSY Forschungsprojekt J\"{u}lich/Bochum,
the Russian Foundation for Fundamental Research
(grants No.\ 03-02-16816 and 03-02-04022),
the INTAS-CALL 2000 N 587.
I am indebted to the organizers of the Conference
for partial financial support.


\newcommand{\noopsort}[1]{} \newcommand{\printfirst}[2]{#1}
  \newcommand{\singleletter}[1]{#1} \newcommand{\switchargs}[2]{#2#1}

\end{document}